%% file: paper.tex
\pdfoutput=1
\documentclass[journal,usenames,dvipsnames]{vgtc}                
\ifpdf
  \pdfoutput=1\relax                   
  \usepackage{graphicx}                
  \DeclareGraphicsExtensions{.pdf,.png,.jpg,.jpeg} 
\else
  \usepackage{graphicx}                
  \DeclareGraphicsExtensions{.eps}     
\fi
\graphicspath{{figures/}{pictures/}{images/}{figs/}{./}} 

\usepackage{microtype}                 
\PassOptionsToPackage{warn}{textcomp}  
\usepackage{textcomp}                  
\usepackage{mathptmx}                  
\usepackage{times}                     
\usepackage{cite}                      
\usepackage{tabu}                      
\usepackage{booktabs}                  

\usepackage{balance}
\usepackage{xcolor}
\usepackage{color}
\usepackage{xspace}
\usepackage{mathptmx}
\usepackage{graphicx}
\usepackage{times}
\usepackage{amsmath}
\usepackage{amssymb}

\usepackage{amsthm}
\usepackage{subfigure}
\usepackage{boxedminipage}
\usepackage{algorithmic}
\usepackage[algoruled]{algorithm2e}
\usepackage{comment}
\usepackage{url}
\usepackage{enumitem}
\usepackage{xfrac}
\usepackage{tabularx}
\usepackage{multirow}
\usepackage{array}
\newcolumntype{M}{>{\centering\arraybackslash}m{\dimexpr.5\linewidth-1.5\tabcolsep}}
\usepackage{bigstrut}
\usepackage[usestackEOL]{stackengine}
\usepackage{bibunits}
\usepackage{wrapfig}
\usepackage{booktabs} 
\usepackage{quotes}
\usepackage{float}
\usepackage{mathrsfs} 
\usepackage{paralist}
\usepackage{longtable,booktabs}
\usepackage{soul}
\usepackage{makecell}
\usepackage{setspace}
\setstcolor{red}

\preprinttext{Online Submission: 1070}

\input{macros}

\onlineid{1198}

\vgtccategory{Research}
\vgtcpapertype{Application}
 
\title{Motion Browser: Visualizing and Understanding Complex Upper Limb Movement Under Obstetrical Brachial Plexus Injuries}

\author{
    Gromit Yeuk-Yin Chan, Luis Gustavo Nonato, \textit{Member, IEEE}, Alice Chu, \\
 Preeti Raghavan, Viswanath Aluru, Cl\'{a}udio T. Silva, \textit{Fellow, IEEE}}

\authorfooter{
    \vspace{-3mm}
Gromit Yeuk-Yin Chan and Cl\'{a}udio T. Silva are with New York University. 
E-mail: \href{mailto:gromit.chan@nyu.edu}{gromit.chan},\href{mailto:csilva@nyu.edu}{csilva}@nyu.edu.
Luis Gustavo Nonato is with University of S{\~a}o Paulo.
E-mail: \href{mailto:gnonato@icmc.usp.br}{gnonato@icmc.usp.br}.
Alice Chu is with Rutgers New Jersey Medical School.
E-mail: \href{mailto:chual@njms.rutgers.edu}{chual@njms.rutgers.edu}.
Preeti Raghavan and Viswanath Aluru are with NYU Langone Medical Center.
E-mail: \href{mailto:Viswanath.Aluru@nyumc.org}{Viswanath.Aluru},\href{mailto:Preeti.Raghavan@nyumc.org}{Preeti.Raghavan}@nyumc.org.

}

\shortauthortitle{Chan \MakeLowercase{\textit{et al.}}: }

\abstract{\input{abstract}}

\keywords{Medical Data Visualization, Visual Analytics Application, Time Series Data, Multimodal Data, Brachial Plexus Injuries}

\teaser{
  \centering
  \vspace{-1mm}
  \includegraphics[width=\linewidth]{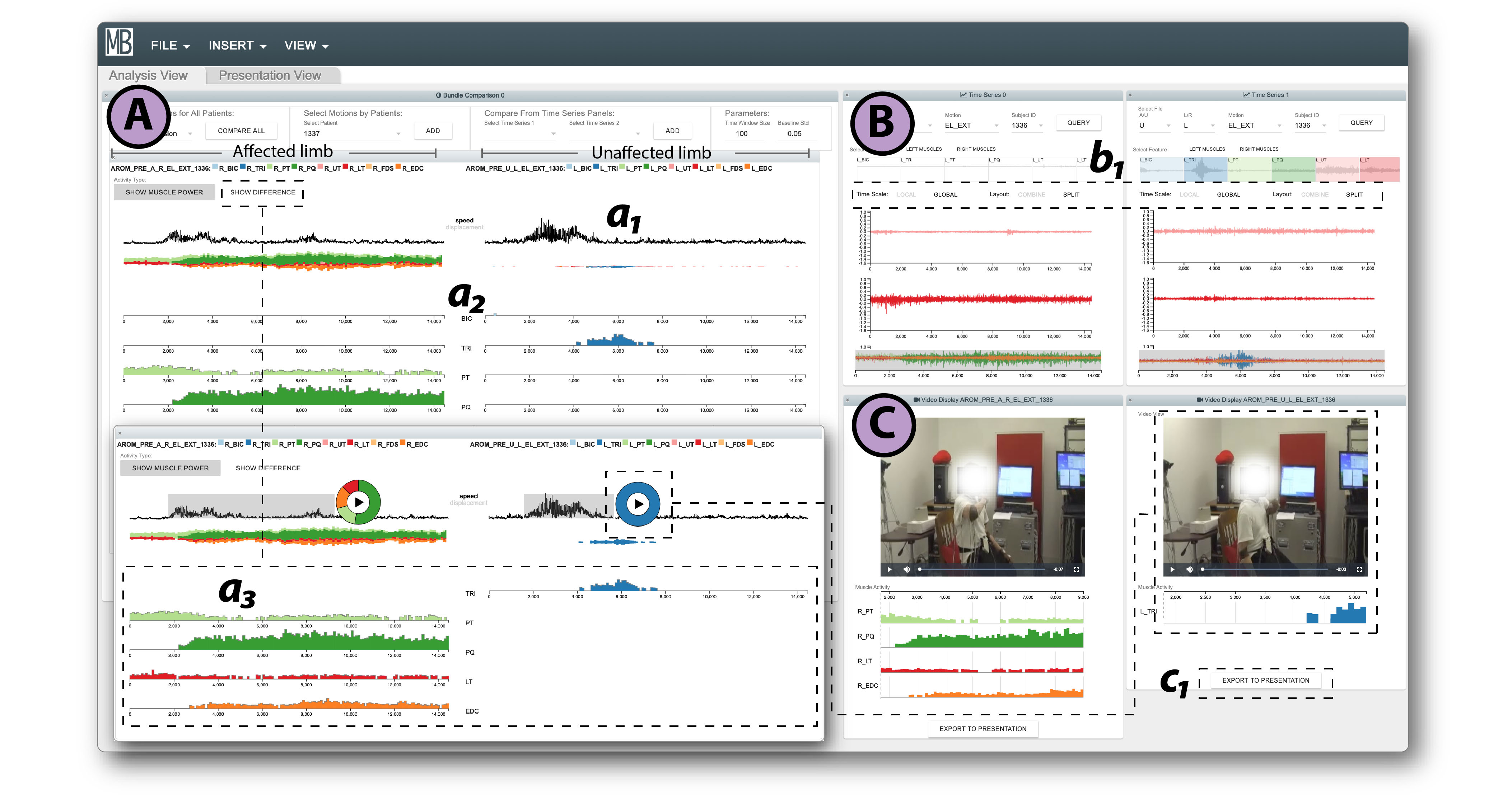}
  \vspace{-8mm}
  \caption{\systemname interface showing how to analyze \revise{patients' limb muscles and movement with data collected from muscle sensors, motion sensors, and video recordings.} 
  \clabel{\textbf{A}} Muscle Bundle Comparison View displays the muscle signals of affected and unaffected limbs side by side.
\revise{Statistics from motion sensors} ($\textbf{a}_\textbf{1}$) and
  stacked muscle activities ($\textbf{a}_\textbf{2}$) are shown.
  Visual highlighting technique allows the extraction of the relatively stronger muscle activities on both sides ($\textbf{a}_\textbf{3}$).
  \clabel{\textbf{B}} Time Series View on raw muscle EMG signals. 
  Each view visualizes the signals from an individual motion 
  and users can align the x- and y-axis of all views ($\textbf{b}_\textbf{1}$).
  \clabel{\textbf{C}} Video Inspection View displays the \revise{cut scenes and filtered signals from \clabel{\textbf{A}}}
  and allows the export to the presentation slide show ($\textbf{c}_\textbf{1}$).\looseness=-1
  }
  \label{fig:teaser}
}

\nocopyrightspace

\vgtcinsertpkg

\begin{document}

\maketitle

\input{introduction}
\input{relwork}
\input{backgroud}

\input{model}
\input{tasks}
\input{desiderata}
\input{system}
\input{use-case}
\input{conc}

\acknowledgments{
%
This work was supported in part by: the Moore-Sloan Data Science Environment at NYU;
NASA; NSF awards CNS-1229185, CCF-1533564, CNS-1544753, CNS-1626098, CNS-1730396, CNS-1828576;
302643/2013-3 CNPq-Brazil and 2016/04391-2 S{\~a}o Paulo Research Foundation (FAPESP) - Brazil. 
C. T. Silva is partially supported by the DARPA MEMEX and D3M programs. 
Any opinions, findings, and conclusions or recommendations expressed in this material are those of the authors and do not necessarily 
reflect the views of DARPA and S{\~a}o Paulo Research Foundation.
}

\begin{spacing}{0.95}
\bibliographystyle{abbrv-doi-narrow}

\clearpage

\bibliography{reference}
\end{spacing}
\end{document}

%% file: macros.tex
\renewcommand{\paragraph}[1]{\vspace{0pt} #1}
\newcommand{\paragraphem}[1]{\vspace{0pt}\noindent \textbf{#1}}

\newcommand{\hidecomment}[1]{}
\newcommand{\savespace}[1]{}

\newcommand{\revise}[1]{{\color{black}#1}}

\newcommand{\systemname}{\textsc{Motion Browser}\xspace}

\newcommand{\etal}{\textit{et al.}\xspace}

\newcommand{\clabel}[1]{\textcircled{\tiny{\textbf{#1}}}}
\newcommand{\cblabel}[1]{\textcircled{\small{#1}}}

\SetKwInput{KwInput}{Input}
\SetKwInput{KwOutput}{Output}

%% file: abstract.tex
The brachial plexus is a complex network of peripheral nerves that enables sensing from and control of the movements of the arms and hand.  
Nowadays, the coordination between the muscles to generate simple movements is still not well understood, 
hindering the knowledge of how to best treat patients with this type of peripheral nerve injury. 
\revise{To acquire enough information for medical data analysis, 
physicians conduct motion analysis assessments with patients} 
to produce a rich dataset of electromyographic signals from multiple muscles 
recorded with joint movements during real-world tasks. 
However, tools for the analysis and visualization of the data in a succinct and interpretable manner are currently not available. 
\revise{Without the ability to integrate, compare, and compute multiple data sources in one platform,
physicians can only compute simple statistical values to describe patient's behavior vaguely,
which limits the possibility to answer clinical questions and generate hypotheses for research.} 
To address this challenge, we have developed \systemname, an interactive visual analytics system 
which provides an efficient framework to extract and compare muscle activity patterns from the patient's limbs 
and coordinated views to help users \revise{analyze muscle signals, motion data, and video information to address different tasks.} 
The system was developed as a result of a collaborative endeavor between computer scientists and orthopedic surgery and rehabilitation physicians. 
We present case studies  
showing physicians can utilize the information displayed to understand how 
individuals coordinate their muscles to initiate appropriate treatment and generate new hypotheses for future research.

%% file: introduction.tex
\section{Introduction}
\label{sec:introduction}

Hands consist of an incredible number of muscles to perform complicated and delicate tasks such as picking flowers without crushing them, 
stringing beads without scattering them, and drinking gracefully from a wine glass without breaking it or spilling its contents. 
The important structures that connect a brain to limb muscles are \textit{brachial plexus}, a complex network of nerves, 
that can be divided into roots, trunks, divisions, 
and cords\cite{sakellariou2014brachial}. These nerves allow brain's control and sensing from the arms to our palms. 
Even though brachial plexus injuries will lead to the disabilities of several muscles, 
the brain will still try to coordinate the remaining functioning parts of the limb to compensate for hand functions. 
As a result, studying the mechanisms of impairment of brachial plexus after injuries and comparing the motions throughout the recovery stages 
can help physicians develop strategies to restore the physical, cognitive, and emotive aspects of hand functions.

\paragraph To study the behavior of muscles from patients under different medical situations, 
physicians need to carry out motion analysis assessments on them with sophisticated setups and protocols. 
They need to attach multiple muscle sensors on patient's hands, forearms and shoulders, and place a camera in front of the patient 
so that he or she can be recorded for the appearances and muscles activities while carrying out several motions. 
As a result, physicians collect a set of heterogeneous datasets that record patients' different kinds of activities, such as elbow extensions, 
shoulder flexions, and wrist rotations, with different sides of limbs completed in different duration. 

\begin{figure}[t]
  \includegraphics[width=\linewidth]{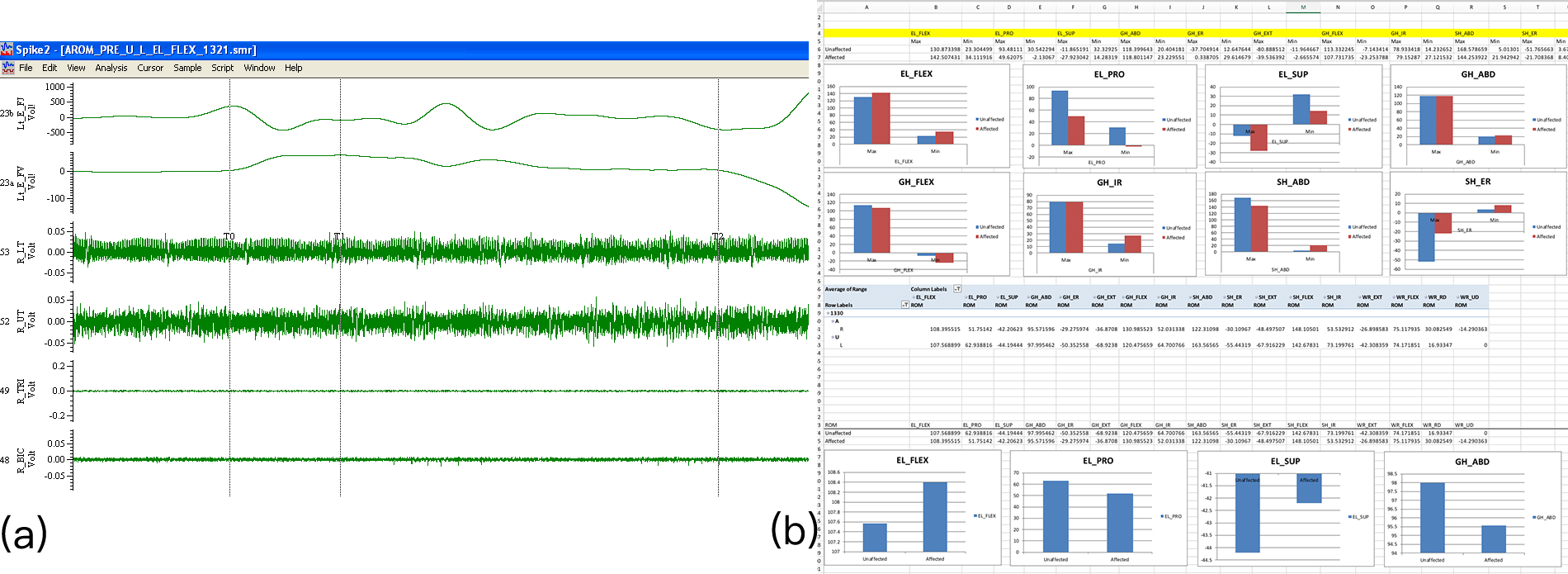}
  \vspace{-5mm}
  \caption{Examples of current tools for analyzing EMG data of muscle activities. 
  (a) Signals of muscle activities and different corresponding physical motions are put together to be inspected in Spike2. 
  (b) Excel Spreadsheet displaying the statistics of muscle activities such as max and min value and maximum range of motions.}
  \label{fig:currentVis}
  \vspace{-5mm}
\end{figure}

\paragraph Currently, physicians rely on visual results from Spike2\cite{smith2003spike2} and Excel to conduct analysis. 
Examples can be seen in \autoref{fig:currentVis}. 
Spike2 is a commercial software that provides signal inspection and analysis techniques such as peak detection and signal decomposition. 
Physicians mainly use it for video editing 
so that the patients' recordings are aligned with the data. 
For Excel, they load the signal data into the cells and plot bar charts with summary statistics such as maximum and minimum values or maximum range of motions. 
Analyzing the data in this way has certain limitations. 
\revise{To begin with, the muscles' activities are hard to be compared among patients.
Each limb is attached with eight sensors so that physicians have to conduct comparative analysis of multiple multivatiate time series.}
Moreover, the muscles can only be quantitatively compared between patient's limbs but not among different patients
\revise{since they depend on the physical strength of patients. Without a standardized metric to normalize the signals among patients,
it is hard to preprocess the data for automated comparison.}
Physicians cannot conclude unless they inspect and compare all patients' behavior one by one in terms of muscle behavior, 
physical outcome, and appearance in the video. 
Furthermore, diagnosis is based on different criteria, such as the differences in muscle coordination between affected and unaffected limbs within the same patient, 
the comparisons of such differences among different patients, or simply the compositions of muscle activities within a single limb.
These require integrations of different attributes and different data abstractions in the dataset on the same platform. 
\revise{Besides, analyzing multiple time series of muscle activities from each patient, 
considered as "muscle bundles"\footnote{Without loss of generality, we refer to multiple time-varying muscle activity signals within one limb as a muscle bundle.}, 
is also a tedious task since physicians have to inspect, analyze, and align multiple muscle signals,
motion sensors, and videos. Without integrating the multimodal data into one platform,
physicians have to conduct the tasks manually with different softwares.}

\paragraph In this circumstance, using a tailored visual analytics approach to address the problems has several advantages. 
First, users can extract patterns in muscle coordination through visual sensemaking actions.
\revise{The muscle signals, which are collected by attaching sensors on the patient's skin surface,
are noisy and ambiguous. 
Muscle contractions come with different shapes of waveforms depending on patients' habits
so that signal processing techniques cannot produce standardized features for comparisons.}
Therefore, providing computed features that highlight the stronger muscles on each limb while allowing users
to steer the final results can balance the validity and efficiency at the same time.
Moreover, provided the complex interplay between a large number of muscle activities,
effective visualization aids the evaluation, reasoning, and communication of the physicians' diagnoses.
Last but not least, a visual analytics system integrating all data sources facilitates a more holistic analysis for the physicians. 

\paragraph To achieve the above-mentioned objectives, we propose \systemname, 
which consists of a novel analytic workflow to compare heterogeneous muscle bundles 
and extract significant muscle activities through semi-automated comparisons with comprehensive visualization techniques, 
as well as an interactive user interface to fulfill the needs physicians to carry out the whole data processing, analytics and communication processes. 
In short, our contributions are as follows:

\begin{compactitem}
\item \revise{An analytics pipeline to visually compare multiple multivariate temporal muscle signals between different limbs.\looseness=-1} 
\item Introduce an interactive visual analytics system to streamline the processes of \revise{inspecting, 
comparing and analyzing the multimodal motion assessment data in an integrated platform.}
\item Present two case studies of physicians using the system
to discover several key symptoms behind how the human brain operates and compromises with impaired nerves, 
in which they result in strategies to speed up patients' recovery processes. 
\end{compactitem} 

%% file: relwork.tex

\section{Related Work}
 
Since our work deals with the challenges of comparing a series of temporal muscle data and analyzing human motion,
we believe that our work can be categorized in the area of motion data visualization, comparative visualization of temporal data
and signal visualization. \looseness=-1

\vspace{-2mm}
\subsection{Motion Data Visualization}
\revise{Many techniques and applications} have been proposed for motion data visualization. 
Readers can refer to a more detailed survey by Bernard \etal \cite{bernard2017approaches}.
Mainly, the motions studied ranged from whole body motions of different entities 
to motions from a part of the body. For visualizing motions of the entire body, FuryExplorer \cite{wilhelm2015furyexplorer} compared the motions 
of horses by visualizing the trajectories projected with PCA from motion sensors attached to different parts of the horses. 
MotionExplorer \cite{bernard2013motionexplorer} visualized motions with hierarchical clustering techniques to distinguish different motions and 
allow experts to cluster motions in a semi-supervised manner.
Krekel \etal visualized the positions of different joints to understand kinematic experiments \cite{krekel2010visual}. 
The system incorporated a 3D skeleton model to help users filter different motions based on the arrangement of the joints.
Keefe \etal proposed a system that used a set of 2D visualizations (i.e. parallel coordinates and line charts) to filter different kinds of 3D motions \cite{krekel2010visual}.
Nguyen \etal abstracted various motion statistics from the knee's volumetric data for interactive analysis \cite{nguyen2016quantitative}.
GestureAnalyzer \cite{jang2014gestureanalyzer} clustered the sequential patterns of human motion data to help users understand different motions,
while MotionFlow \cite{jang2016motionflow} was a continued work that addressed the same challenges through comparative visualization. \looseness=-1

Despite having a similar goal of understanding body behavior, 
\systemname differs from the current literature by analyzing the \revise{complex muscle coordination with the help of motion data.}
Our main goal is to uncover the relationships of muscles when a limb is injured, with the help of visible motions that 
act as verification as well as an indicator for physicians to leverage their domain knowledge to enhance the analysis. \looseness=-1







\vspace{-2mm}
\subsection{Comparative Temporal Data Visualization}
Time series visualization and comparative visualization are areas that earn great attention in the realm of visual analytics.
Readers can refer to \cite{aigner2011visualization,aigner2012comparative,du2017coping} for an in-depth survey of how time series is visualized or processed
for different kinds of goals and abstractions, and \cite{gleicher2018considerations,kehrer2013visualization} for how comparative visualization is considered and designed.
\revise{Both visualization and comparison are tightly integrated for temporal data analysis.} 
For example, the initial motivation of Playfair's invention of line chart was to illustrate comparisons \cite{tufte2001visual}.
When considering the design of temporal data comparison, literature mainly studied different layouts and graphs to satisfy specific tasks \cite{javed2012composite,javed2010graphical}.
The layouts could either be juxtaposition, superposition, or explicit encoding, while graphs could be either stacked, small multiples or overlaid simple graphs. 

Various \revise{techniques and systems} were proposed to compare temporal data. 
Kehrer \etal applied focus+context techniques on climate dataset to present different abstractions at different levels to facilitate comparisons \cite{kehrer2008hypothesis}.
SimilarityExplorer \cite{poco2014similarityexplorer} compared spatiotemporal climate models by separating the models' space and time attributes with multiple views for comparisons.
Temporal Summaries \cite{wang2009temporal} used categorical sequences as the abstraction target to facilitate comparison tasks.
Li \etal used multimodal visualization to filter time series with additional attributes \cite{li2016visual}.\looseness=-1

Electronic signals require much interactivity when it comes to exploratory data analysis (EDA).
Chronolenses\cite{zhao2011exploratory}, Kronominer\cite{zhao2011kronominer}, SignalLens\cite{kincaid2010signallens}, and TimeSlice\cite{zhao2012timeslice} 
proposed interactive systems to address the challenges on volume and variety of visualizing these temporal data in general. 
There were also other data abstraction techniques used on multivariate signals.
Dal Col \etal \cite{dal2017wavelet} and Valdivia \etal \cite{valdivia2015wavelet} abstracted multivariate signals as connected graph based on correlations,
so that wavelet theory could be applied to depict similarity of groups of time segments to generate abstractions for visualization.
Ward \etal extracted signals as N-grams to project the abstraction with dimensional reduction techniques (i.e. PCA) \cite{ward2011visual}. \looseness=-1

Our work's novelty compared with the above work is driven by our physicians' needs of sequentially scanning multiple comparisons, 
meaning that we aim at facilitating the comparisons of comparisons. 
Our comparative visual analytics components have a focus on how to help physicians quickly compare patients' own healthy and affected limbs, 
and then use the results to enable comparisons across different patients.
Also, our work focuses on the visualization of comparisons between different \textit{groups} of multivariate time series, in which our techniques emphasize the clarity on comparisons, 
their extensibility on analysis, and interactivity.

%% file: backgroud.tex
\begin{figure*}
  \includegraphics[width=\linewidth]{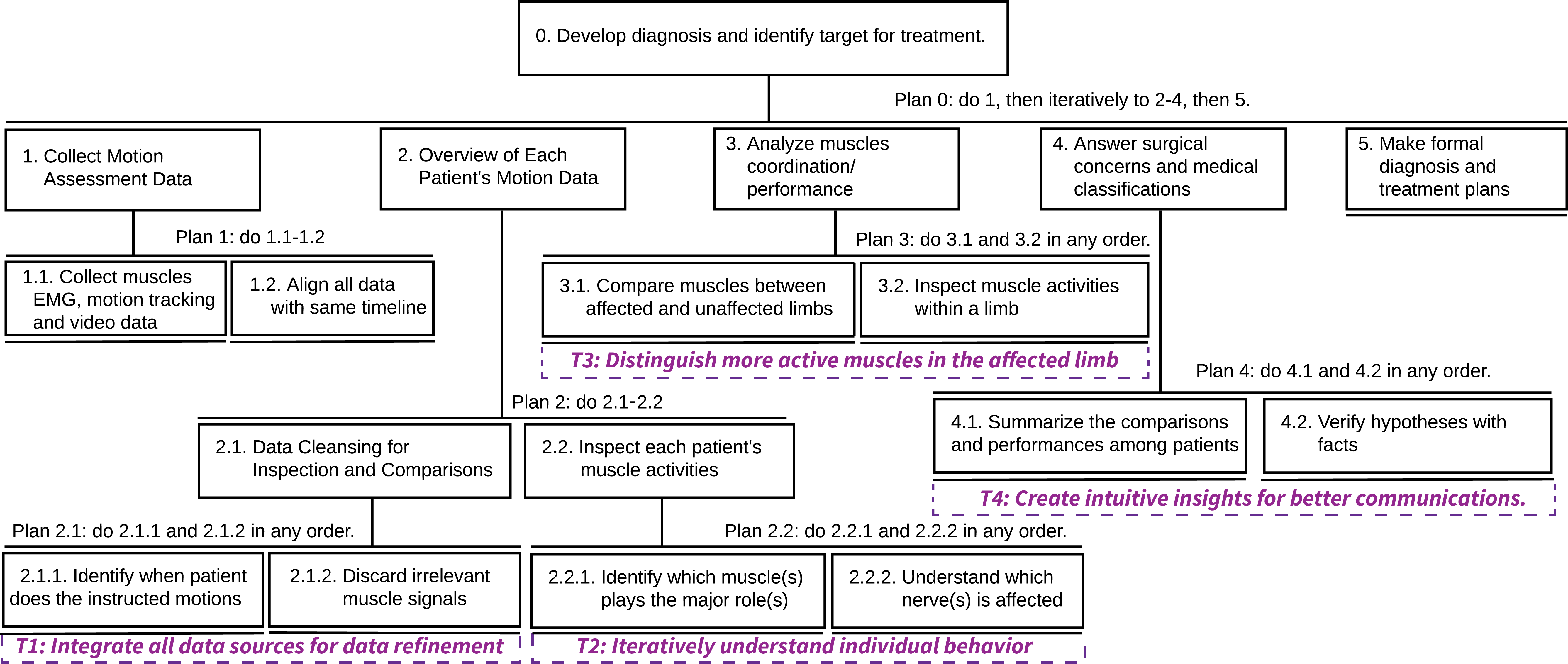}
  \caption{Hierarchical task abstraction of the clinical workflow. 
  Each box represents a task or subtask and each level of hierarchy has a plan. 
  The horizontal line at the bottom of the box means a termination. 
  The highlighted purple text represents the \textbf{\textcolor{Fuchsia}{task abstraction}} derived from the tasks.
  \looseness=-1}
  \vspace{-3mm}
  \label{fig:hta}
\end{figure*}

\section{Domain Problem Characterization}
\label{sec:background} 
The brachial plexus is a set of nerves supply to the upper limb. 
\textbf{Obstetrical Brachial Plexus Birth Palsy} (OBPBP) refers to injury noted in the perinatal period, 
which is around the time of birth, to all or a portion of the brachial plexus. 
There are many competing theories as to the source of dysfunction in OBPBP and as many different modalities of treatment. 
The most often cited theories for continuing dysfunction are residual weakness, muscle co-contraction, glenohumeral dysplasia, 
joint or muscle contracture, and ineffective compensatory technique\cite{anguelova2014extensive,sheffler2012biceps}. 
The challenge is that the treatment for each of these etiologies is different and sometimes contradictory. 
To obtain a logical algorithm of solutions, 
physicians need a more detailed understanding of motion in obstetrical brachial plexus patients, 
especially how various muscles coordinate under different levels of severity or symptoms. 

In general, there are score systems such as \textit{Narakas Classification}\cite{al2009narakas} to assess possible outcomes of patients. 
Yet, they are based on clinical observations and the scores may vary among different clinicians. 
To make treatment plans that do not only base on physical evidence but also muscle performances, 
our physicians initiated the \textbf{Pediatric Upper Extremity Motion Analysis Program} in 2013, 
which modified the equipment for adult’s motion analysis to facilitate pediatric motion analysis. 
In this \textbf{Active Range Of Motion} assessment, 
the patients were connected with 8 sensors that measured activities in terms of 16-channel electromyography (EMG) signals from 8 muscles. 
They were also monitored with 3-dimensional motion analysis and video recordings. 
Using the data, our physicians attempted to acquire insights and generate data facts 
that helped them explain the rationale behind their concluded treatment plans to other clinicians. 
This eventually resulted in the need for a holistic system to improve muscle analysis and data communication.

While our study ultimately aims at helping clinicians address medical challenges and improve clinical workflow, 
the process of creating \systemname is \revise{an example that helps provide evidence} 
in creating a meaningful application that \revise{applies interactive visualization to solve a medical problem}. 
To provide a design study that balances both the values of the medical domain and visual analytics, 
we shaped our development under the guidelines of Multi-dimensional In-depth Long-term Case studies (MILCs)\cite{shneiderman2006strategies}. 
In short, we emphasize two important aspects to maximize the understanding of the visualization design process. \looseness=-1

\textbf{Breaking Domain Experts' Goals to Hierarchies.} 
Our physician experts hope to understand how muscles coordinate in patients with brachial plexus injuries, 
and they applied a set of baseline procedures to conduct the analysis. 
Therefore, it is important to get both the high-level goals and low-level details to address experts' key needs of interactive visualization for our application. 
To do this, we paid clinical visits to conduct contextual inquiry \cite{beyer1995apprenticing} interviews
to understand how they conduct hands-on diagnoses on a total of 8 patients using the data they collected from the motion analysis. 
We formulate the whole process exclusively and exhaustively such that it becomes a Hierarchical Task Analysis (HTA)\cite{shepherd1998hta,zhang2018idmvis}. 
In HTA, actions from the domain experts become a set of tasks and subtasks. 
Each task has a goal and a plan. 
HTA allows us to identify the tasks that can potentially benefit the most from interactive visualization, 
such that it helps justify our system requirement. \looseness=-1

\textbf{Iterative Process of Design.} 
Overall the project took about 12 months. Our visualization researchers worked tightly with two Orthopedics and Rehabilitation physicians 
in shaping the scope, identifying the objectives and iterating different system designs. 
One physician had regular clinical checkups with the 8 patients and 
another physician-scientist was responsible for conducting the motion assessments.
We first replicated some baseline features such as time series plots of EMG signals, 
and then the physicians requested more features for generating results and different forms of communications from the data. 
We, therefore, held regular (bi-weekly) meetings to evaluate the available features and then introduce different designs and algorithms for the physicians to evaluate.  
Our physicians also occasionally introduced the tool to other researchers and collected their feedback.

%% file: model.tex

\section{Data Abstraction}
\label{sec:data}

The \textit{Active Range Of Motion} assessment mentioned in \autoref{sec:background} generates the AROM dataset to initiate our analysis. 
Each motion consists of EMG signals of 8 muscles from each side of the upper limb, allowing physicians to acquire muscle activities throughout patient's forearms, 
elbows, hands, and fingers. 
Alongside this, a patient's limb motion is also tracked with motion trackers and video recordings. 
Overall, the EMG signals play the main role in comparing the patients' muscle coordination, 
while the motion tracking data provides the context of patients' performances. 
The videos act as verification and direct inspection of the whole motion.
The whole dataset is summarized in \autoref{table:dataAbstraction}. \looseness=-1

\begin{table}[H]
\centering
\caption{Data Abstraction Summary}
\vspace{-2mm}
\label{table:dataAbstraction}
\begin{tabular}{lll}
\hline
Key Attributes   & Data Abstraction         & Objectives          \\ \hline
Muscle Activity  & Quantitative (EMG Signals)   & Comparisons        \\
Motion Tracking  & Quantitative (Positions) & Context \\
Video Monitoring & Images                   & Verification       \\ \hline
\end{tabular}%
\end{table}

\noindent\textbf{Muscle Activity Signals.} The EMG signals play a central role in helping physicians clarify the coordination of muscles throughout patients' motions.
While the signals capture the muscle activity with their amplitudes, 
they are hard to read and compare in the raw format. 
To first facilitate the inspection of EMG signals, 
they are usually transformed with root-mean-square (RMS) envelope, which is calculated using a time-windowed RMS function:
\begin{equation}
  RMS=(\frac{1}{S} \sum_{1}^{S}f^{2}(s))^{\frac{1}{2}}
  \label{eq:rms}
\end{equation}

The RMS value represents the power of the signal, which correlates with the degree of muscle activity and is always positive. 
As the raw EMG signals are oscillating and produce more clutters for visualization, 
turning them into RMS values produces a polarized waveform that is more easily analyzable.

However, there are two challenges from the EMG signals that cannot be solved solely by automatic computation.
Firstly, the EMG signals are only normalized within the same person. 
It means that the amplitude of muscle signals can only be compared between a patient's left and right limb but not among different patients. 
Physicians thus use them first to find out which muscles in the affected limb are stronger than the unaffected limb to deduce the compensatory muscles,
then compare the presence of stronger muscles among different patients.
We address the comparisons of patients by the comparisons within their limbs as the \textit{scalability} challenges of our visual analysis.
Secondly, we have to deal with the \textit{noisiness} of muscle signals. 
As the EMG signals are collected from the skin, there might be an ambiguity of the resulting power.
We, therefore, have to ensure human-in-the-loop throughout the analysis. \looseness=-1

\noindent\textbf{Data from motion tracking.} The 3-dimensional motion analysis tracks the patient's limbs' coordinates in x, y and z directions. 
These attributes enable the derivation of speed and acceleration of limb motions, 
which allow physicians to locate a finer scope of motion and compare the overall performance without inspecting the video.

\noindent\textbf{Video Monitoring of Motions.} The video monitoring of the whole motion provides a full picture of the patient's performance. 
Each cut scene lets physicians verify their findings from the analysis of muscle signals. \looseness=-1

%% file: tasks.tex
\section{Task and Requirement Analysis}
\label{sec:tasks}   
\subsection{Hierarchical Task Analysis of Clinician Workflow}
Mentioned in \autoref{sec:background}, we break down the procedures of conducting an analytic workflow from the domain perspective in a Hierarchical Task Analysis. 
In the clinical diagnosis, the physician experts want to decide \textit{which muscles, joints or nerves should be targeted for treatment} (\textit{Task 0}). 
To prepare for the analysis, they first modify the equipment to conduct a series of motion analysis (\textit{Task 1}). 
The motion data, including muscle activities, motion data, and video recordings, are recorded from both the affected and unaffected limbs 
so that the muscle EMG signals can be normalized and compared between both sides (\textit{Task 1.1}). 
Then, they use video editors to trim the videos such that all the data attributes are temporally aligned (\textit{Task 1.2}). 

After the physicians acquire the data, they inspect each patient's limb motion (\textit{Task 2}). 
Throughout the motion analysis, there may exist irrelevant information related to the specific motion. 
For example, physicians do not need to know the fingers' muscle activities when they analyze a patient's shoulder rotation, 
or the patient only spend 5 seconds to finish the action throughout the 15 seconds recording. 
As a result, the objective of inspecting everything is to extract the relevant information from the motion assessment (\textit{Task 2.1}). 
When physicians locate the useful portion of data (\textit{Task 2.1.1}) and remove the irrelevant parts (\textit{Task 2.1.2}), 
they will be able to acquire useful overview (\textit{Task 2.2}). 
Browsing all the information regarding muscle activities, motion and videos together,
physicians try to acquire a general impression of the performance such as which muscle(s) play(s) the main role (\textit{Task 2.2.1}) 
or which of them is/are injured (\textit{Task 2.2.2}). 

At this stage, physicians already grasp a basic understanding of how the patient behaves, 
so now they can analyze how is the coordination of muscles compromised for the affected limb (\textit{Task 3}). 
This is done by comparing the difference between the patient's limbs. 
For example, physicians inspect which muscles in the affected limb react stronger, or vice versa (\textit{Task 3.1}). 
Once physicians identify which muscles play important roles in each limb, they can separate and highlight them (\textit{Task 3.2}). 
For example, if the affected limb is identified to have higher activities of biceps and triceps compared with the healthy limb, 
then the physicians understand these muscles play an active role in compensating the movement. 
Such facts allow them to compare the results among different patients. 
By comparing the comparisons, physicians will be able to answers questions like 
``Does trapezius muscle play an active for patients with brachial plexus injuries?'' \looseness=-1

Therefore, when the physicians finish analyzing each performance, 
they need to put all the patients' results on a single page to answer surgical concerns (\textit{Task 4}).
Usually, they first classify patients into different groups based on their performances,
such as ``high functioning'' and ``low functioning'' groups,  
and then summarize the muscle activities in each group (\textit{Task 4.1}). 
The summaries are in the form of proportions to let physicians understand which muscles are prevalent as compensatory muscles among the patients.
For example, if biceps and triceps are the two muscles that behave stronger in the affected limb on one patient, 
the physicians would like to know their ratios among other patients to see if these muscles are common priorities to compensate the motions.
Based on these, 
they can generate hypotheses and verify their reasoning with video recordings at the same time (\textit{Task 4.2}).
Note that a diagnosis made with such analysis is empirical and based on observations 
so that physicians need to pack all of the findings in a presentable format that includes charts and video screenshots and send the report to other colleagues. 
The presentation mainly acts as evidence that helps physicians explain more in-depth medical knowledge phenomena. 
Once the consensus is reached among several physicians, then they will land a more formal diagnosis and treatment plan (\textit{Task 5}).

\subsection{Hierarchical Task Abstraction}
\revise{Based on the hierarchical task analysis, we translate the domain tasks into abstract forms to generate tasks in the visual analytics domain.}
Based on our interview with the domain experts, 
we noticed that while the task abstractions of many design studies aim at exploratory data analysis \cite{lam2018bridging}, 
our main challenge is to combine domain knowledge and computation power to make the analysis more \textit{streamlined and efficient}. 
Instead of using multiple software to glue the heterogeneous data together and just solely put the signals side by side for comparison, 
our experts want to have minimal interactions so that they can go through each patient one by one faster. 
Therefore, we follow several iterations in the \textit{nested model} \cite{munzner2009nested} and loop through the four nested layers to refine our requirements
from the task abstractions (Fig.\ref{fig:hta}).  The high-level task abstractions apart 
from the operation contexts (i.e. \textit{Task 1 and 4}) that our physicians want to achieve from the data analysis are summarized as \textbf{T1-4} in Fig.\ref{fig:hta}.

%% file: desiderata.tex
\vspace{-2mm}
\subsection{Design Requirement}
\label{sec:desiderata}
We identified design requirements based on the informal qualitative
interview with physicians and task abstractions discussed above.

\begin{compactitem}
\item[\textbf{R1}] \textbf{Align heterogeneous data sources.} 
Physicians need to analyze information from muscle and motion sensors and video recordings 
\revise{to compare the muscles within a patient's limb or between his affected and unaffected limb (\textbf{T.1}).} 
\revise{To speed up the inspection and exploration, the display should align the information by time,
patient, and muscles.}

\item[\textbf{R2}] \revise{\textbf{Display and analyze individual patients' performance in one view grouped by each patient.}}
\revise{Our experts indicate that it is a clinical practice that they need to analyze patient's data one by one before summarizing their behavior as a whole.
Therefore, the system should display information and analysis of each patient in an individual view, while showing all information in the same window,
so that physicians can explore, scroll, and analyze the patients effectively (\textbf{T.2}).}

\item[\textbf{R3}] \textbf{Enable efficient comparative analysis.} 
It will be time-consuming and cognitively overwhelming to extract stronger muscles on each limb by manual selection (\textbf{T.3}). 
Physicians need an efficient mechanism to compare muscle coordination between each patient's limbs,
\revise{then use the comparisons to understand the difference among all patients.}
\revise{While the definition of stronger muscles is subtle and depends much on observation,
some similar muscle activities, such as similar waveforms and magnitudes, should be filtered away at the beginning.}

\item[\textbf{R4}] \revise{\textbf{Export the clinical analysis for presentation.}} 
\revise{After finishing the data analysis (\textbf{T.2}), 
experts need to report their findings on the comparison among patients for discussion and verification (\textbf{T.4}). 
An easily communicated visualization is preferred to showcase their results to other colleagues.\looseness=-1} 

\end{compactitem}

%% file: system.tex

\section{Visual Analysis of Muscle Activity Comparison}
\label{sec:bundle}
In this section, we present the main visual analytics component of the system, the muscle bundle comparison chart, 
for comparing the muscle coordination between a patient's affected and unaffected limb.
\revise{This component contains the main analytics pipeline in analyzing the muscle signals, while the other views in the next section are responsible for information and presentation.}
Because there exists a tradeoff between visual clarity and interactivity, 
we introduce a visual highlighting technique using an entropy computational metric to quickly extract meaningful comparisons.

\begin{figure}[ht]
	\includegraphics[width=\linewidth]{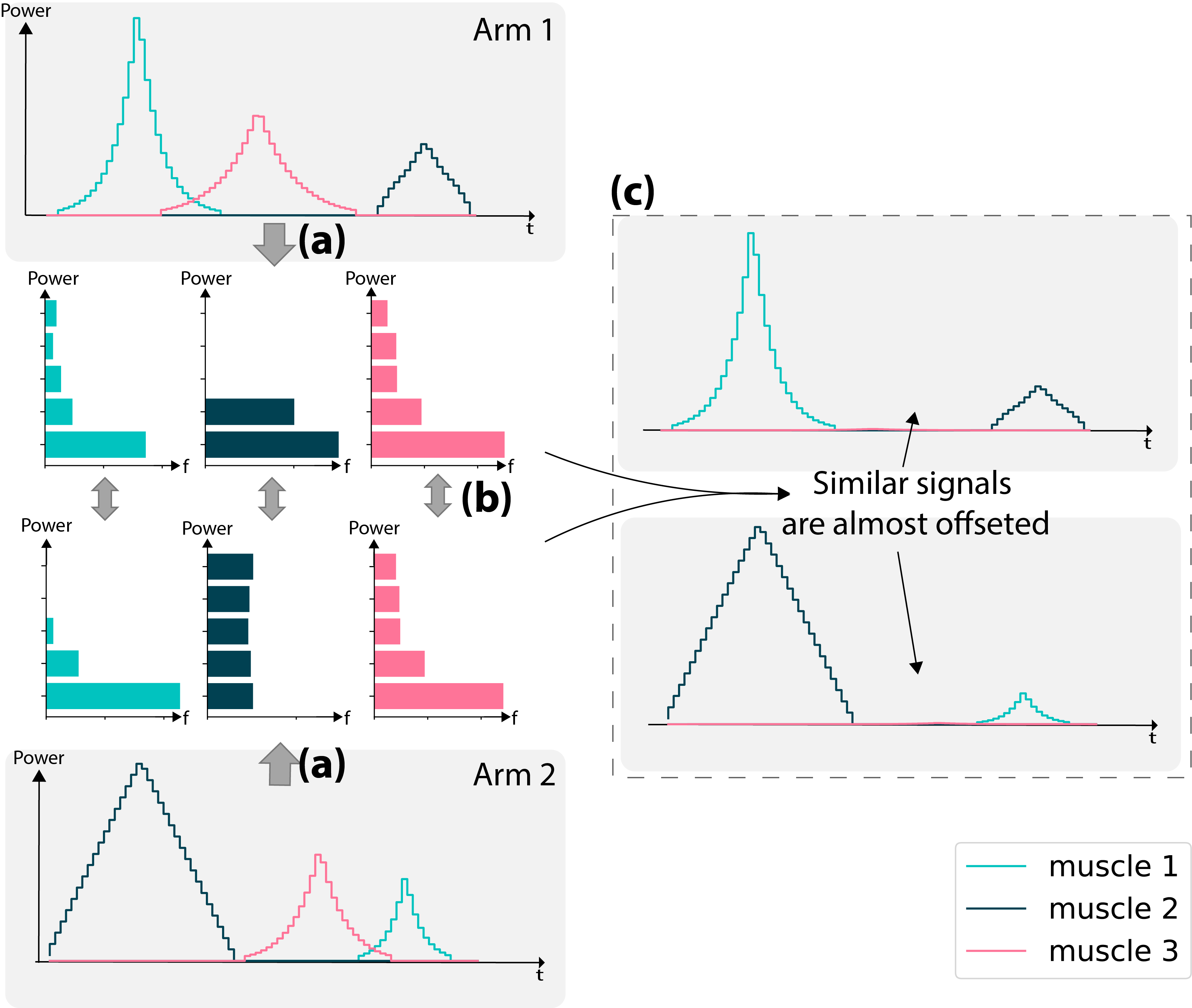}
	\caption{Entropy-based visual highlighting: (a) Transform the signals of two limbs to histograms of value, then (b) compare the ordinal distributions of the histogram. 
	(c) The similar and smaller muscle signals are reduced 
	so that significant signals eventually stand out. \looseness=-1}
    \label{fig:illustration}
    \vspace{-3mm}
\end{figure}

\subsection{Visual Design}
\revise{To begin with, we first address the requirement of displaying all sensors of each patient in one view (\textbf{R2}).}
A bundle comparison chart displays the motion assessment information of patient's both limbs (\autoref{fig:teaser}\clabel{\textbf{A}}, \revise{\autoref{fig:illustration}}) side by side. 
The left-hand side shows the data from the affected limb
and the right-hand side shows the data from the unaffected limb. 
The line chart in the view (\revise{\autoref{fig:illustration}}) displays the quantitative outcome of the motion (i.e. speed or displacement).
When users brush on the line chart, the proportion of muscle activities in the brushed region will be shown as a donut chart (\autoref{fig:teaser}, \revise{\autoref{fig:case2}}).
The bar charts show the muscle activity signals. 
The signals are either stacked or 
arranged in small multiples vertically (\autoref{fig:teaser}($a_1$), \revise{\autoref{fig:illustration}}). 
Color encodes the muscles and the same muscles from both limbs contain the same color as well.
We use the eight qualitative color scale from ColorBrewer \cite{harrower2003colorbrewer} in a way that 
two bluish colors \raisebox{-.2\height}{\includegraphics[scale=0.18]{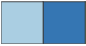}} correspond to two pushing muscles, 
greenish colors \raisebox{-.2\height}{\includegraphics[scale=0.18]{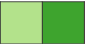}} correspond to forearm muscles, 
reddish colors \raisebox{-.2\height}{\includegraphics[scale=0.18]{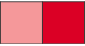}} correspond to back muscles and 
yellowish colors \raisebox{-.2\height}{\includegraphics[scale=0.18]{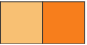}} correspond to finger muscles. 
In an aligned way, the activities from the same muscles in both limbs are also plotted together vertically.
Lastly, every information is also aligned with time horizontally. 

\subsection{Design Alternatives}
We experiment with various visual techniques by iterating several layouts and approaches with the physicians.
Visual comparison of a series ordered data can be accomplished by juxtaposition, 
superposition or explicit encoding \cite{gleicher2018considerations}. 
Explicit encoding does not apply to our problem because the relationships between different muscles are hard to calculate, which is
one of the reasons for us to develop a comprehensive system to address the challenge. Superposition, either by placing 16 muscles together
or 2 compared muscles in each small multiples, causes the alignment of motion information and muscle data to become obscure. 
As the physicians want to identify which muscles
contribute to rapid movement or limb positioning, 
they prefer inspecting every information in a separate chart for better clarity (\textbf{R2}).

The next consideration is whether we should stack each muscle activity on one chart instead of placing them in small multiples.
We first tried stacking the muscle activities from one limb together to squeeze more muscle comparison views
on one screen. While the physicians appreciated the efficiency of scrolling through patients in such layout, they raised a problem of perceiving
small but important muscle signals in a limb. Sometimes if a muscle emits a relatively weaker signal compared with muscles inside the limb but is indeed stronger than
the same muscle across the other limb, it is still treated as important. Such inconsistency of scaling makes small multiples a better choice in 
comparing the difference \cite{javed2010graphical}. Thus, in our current version, we include both layouts for the physicians to inspect both forms. \looseness=-1

Moreover, we consider the alternatives of donut chart when brushing the line chart. 
While the physicians were keen on seeing the percentage and using it as a button to play the video at the same time, 
our first version encoded the donut chart's radius with the total amplitude of muscle signals. 
It turned out not to be a good idea since when the chart became too small, the physicians could not inspect anything meaningful. 
Nonetheless, the amplitude information can already be perceived in the stacked charts. 
Therefore, we do not encode any more details on the donut chart except the proportion of muscle signals. \looseness=-1

\begin{figure}[tb]
	\includegraphics[width=\linewidth]{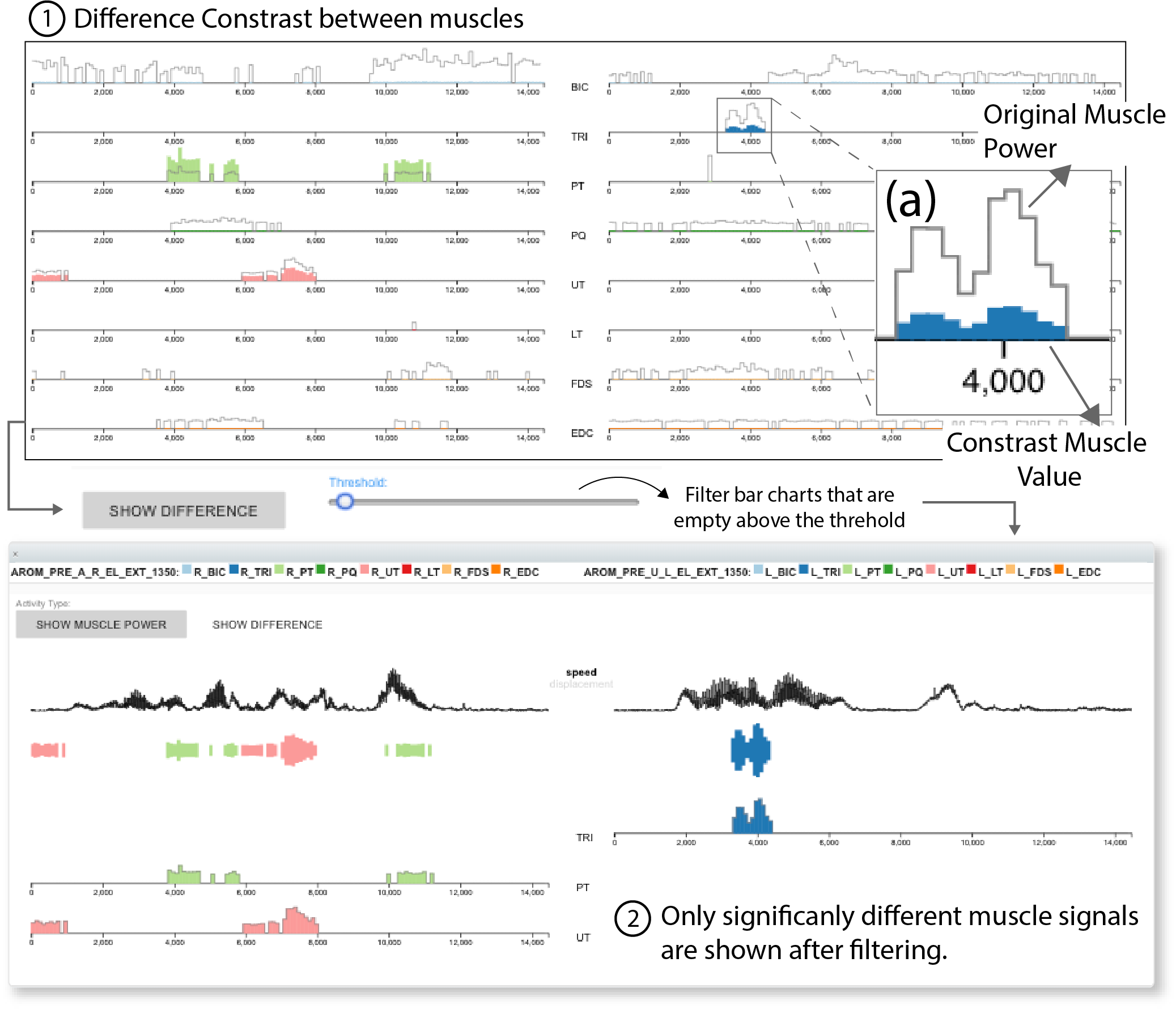}
	\caption{Illustration of how users can discover the significant muscles on both limbs. 
            \clabel{1} When users apply the highlighting option, the highlight of muscle activities in \autoref{fig:teaser}\clabel{\textbf{A}} 
            change based on their relative significance.
			(a) The original power is encoded with unfilled lines.
            \clabel{2} Sliding through the threshold filter, the charts without any values left will collapse, 
            reducing the number of muscles shown at the end. \looseness=-1
            }
    \vspace{-3mm}
	\label{fig:bundleComparison}
  \end{figure}
\subsection{Entropy Based Visual Highlighting}
The separation of all data in the view aims at helping the physicians examine all sources of quantitative data in a consistent and aligned time scale (\textbf{R1}).
However, now if everything is separated, there exists a challenge in interactivity,
as users need to click on each of the bar charts one by one to remove the muscles that are similar between the limbs. 
If none of the charts are removed, the number of visuals will accumulate when users scan each patient's bundle comparison views sequentially.
Thus, either remove or not remove will overload users' cognitive ability easily.
Therefore, we investigate the possibilities for physicians to inspect the information clearly while efficiently extracting the more significant muscle activities on both sides of the limbs (\textbf{R3}).
We introduce a method that can offset similar signals and reduce the presence of smaller signals
so that the significant signals will eventually stand out.\looseness=-1

\autoref{fig:illustration} illustrates the method. First, for each muscle activity, we put all the values into a histogram. 
Then we calculate the difference between the histograms of the same muscles in both limbs, using Kullback-Leibler divergence:
\begin{equation}
  D_{KL}(Q||P)=\sum_{i} Q(i)ln(\frac{Q(i)}{P(i)})
  \label{eq:kld}
\end{equation}
$P$ and $Q$ are the distributions of values in each muscle's activity.
This function measures how one distribution deviates from another reference distribution
and is used as highlights in several situations of visualization \cite{correll2017surprise,hullman2013contextifier}.
In the example here, when two distributions are similar to each other, 
each log value in the summation will become very close to zero, leading to a low divergence. 
On the other hand, there are great differences between each bucket, the divergence becomes high.
We also calibrate the signals with the skewness\cite{kokoska1999crc}, 
such that if the distribution is more left-skewed (i.e. more small values) the value will decrease.
The histogram is constructed by K Means clustering. For simplicity, we use the elbow method to decide the value of K.
In this way, we can visualize a larger number of muscle signals and still quickly identify the significant muscle activities. 
\begin{figure}[t]
	\includegraphics[width=\linewidth]{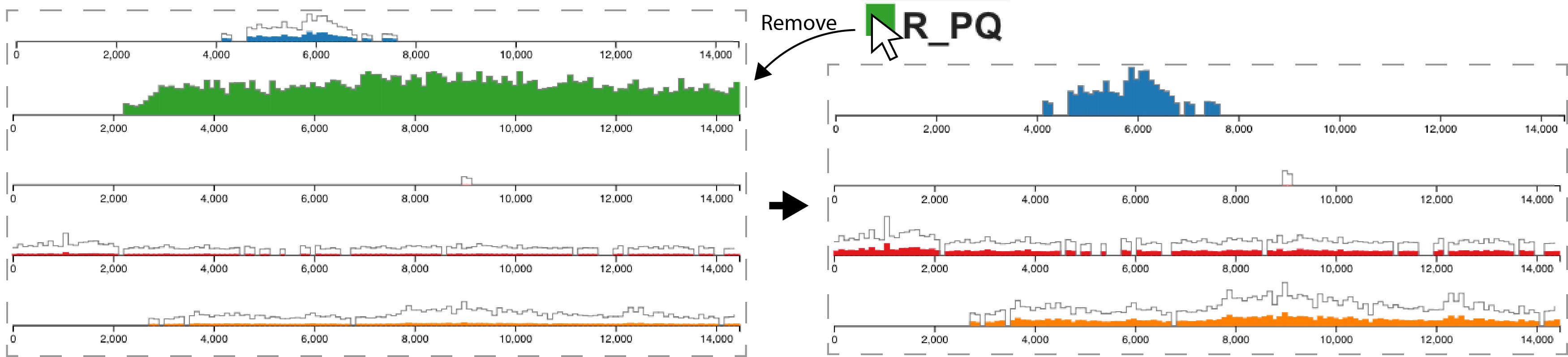}
	\caption{
    When users remove a muscle, 
	the corresponding bar charts will collapse and the remaining ones may rescale to become more visible. \looseness=-1}
    \label{fig:removeMuscle}
    \vspace{-3mm}
  \end{figure}
\vspace{-2mm}
\subsection{Interactive Identification of Important Muscles}
We introduce a mechanism of interaction to efficiently identify the significant muscle signals with the help of our visual highlighting method.
When users click to apply the muscle highlighting, they can inspect the highlighted results illustrated in \autoref{fig:bundleComparison}\cblabel{1}. 
The results are encoded in colored area charts and the strokes in the chart encode the original power (\autoref{fig:bundleComparison}(a)) 
so that users do not need to switch back and forth to recall the original muscle signals.
Our method of grasping the significant muscle signals can be summarized as \textbf{sliding} and \textbf{collapsing}. 
When users slide and filter by highlighted values,
the bar charts that become empty will disappear. If bar charts on both sides both disappear, the whole space will collapse. 
As a result, the final display of muscle activities will be distilled to only the muscle signals 
that are significantly greater than the opposite limb (\autoref{fig:bundleComparison}\cblabel{2}). 
On the other hand, users can manually remove the bar charts by checking the legend. 
In both ways, the remaining bar charts will resize such that the muscles may become more visible 
due to the rescaling of the y-axis (\autoref{fig:removeMuscle}).\looseness=-1

The sliding mechanism engages users to opt for a clear comparison with a few amounts of interactions needed
so that they can obtain minimal operations to analyze each patient sequentially (\textbf{R3}).

\section{The \systemname System}
\label{sec:system}
\subsection{Analysis View}
To support users analyzing each's behavior, our system consists of a query panel to select the dataset, the muscle bundle comparison view, 
and a video inspection view. 
These displays follow a hierarchical relationship in a way that follows the visual analytics mantra 
``overview first, zoom and filter, then details-on-demand'' \cite{shneiderman1996eyes}, from the perspective of analyzing an individual patient.
\revise{A general overview of the workflow is as follows: first, the user obtains and analyzes each patient's data in the bundle comparison view in \autoref{sec:bundle}, then the processed details are exported to the video view for comparison among all patients. After that, the findings are shown in the presentation view.\looseness=-1}


\textbf{Time Series View}
Users can inspect raw muscle EMG signals in the time series view (\autoref{fig:teaser}\clabel{\textbf{B}}). 
They can query with different file types to select the motion of a patient's limb.
Also, the scales across different time series views can be aligned so that different time series panels can be compared with each other.
This view acts as a time series editor when users encounter an erroneous motion (i.e. unreasonably long recordings or abnormal muscle signals).
Users can visually refine the timeline and muscle selection (\textbf{R1}) and import the refined results to the bundle comparison view. \looseness=-1


\textbf{Video View} 
After users click the play button in the selected time intervals (\autoref{fig:teaser}$\textbf{a}_{\textbf{3}}$), 
a new window will appear to show the selected muscle activities under the video. 
All of the information is aligned with a line to synchronize the video time frame and charts (\autoref{fig:teaser}\clabel{\textbf{C}}).
Such encoding allows a compact integration of all information and insights obtained aligned with video evidence (\textbf{R4}).
Users thus can verify their findings and derive reasons between muscle coordination and physical outcome.\looseness=-1

\begin{figure}[!htb]
	\includegraphics[width=\linewidth]{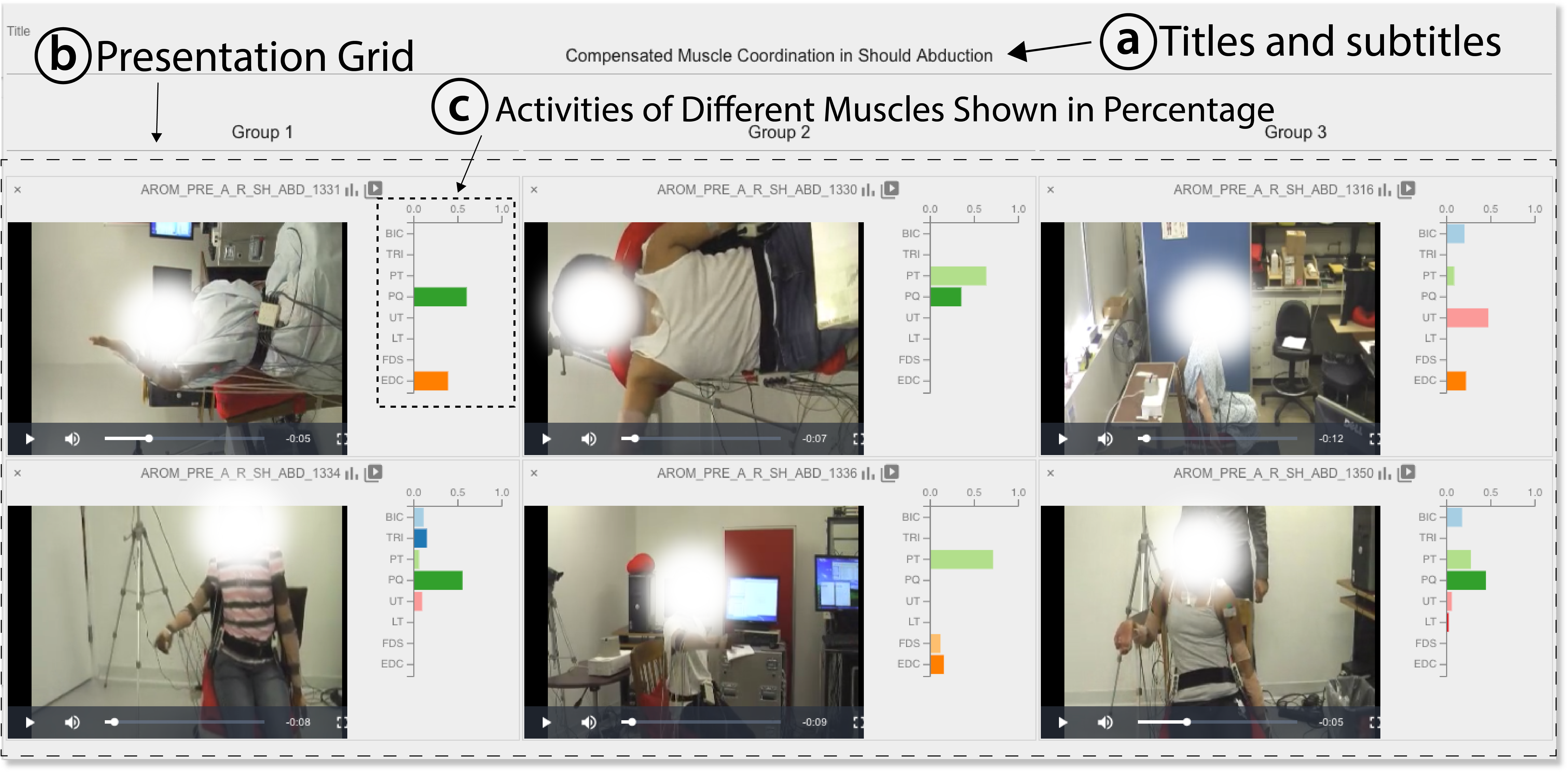}
	\vspace{-6mm}
	\caption{Presentation View. Users can create a presentation for their analyzes by \clabel{a} annotating with titles and subtitles; 
	\clabel{b} arranging each insight from video views in the presentation spreadsheet, and;
	\clabel{c} presenting the muscle activity insights on each limb in percentages.
	}
	\label{fig:presentationView}
	\vspace{-3mm}
  \end{figure}

\subsection{Presentation View}
After users finish inspecting the \revise{results} they obtained in the video view, they can export these video snippets to the presentation view (\autoref{fig:presentationView}).
This view is a grid layout that allows users to align the insights from the analysis to compare different patients and add annotations to explain the findings (\textbf{R4}).
Each patient's muscle activities are shown as a percentage since the percentage of muscles used can be compared across different patients.
For example, experts can understand a patient uses more biceps to compensate for his shoulder movement than other patients by showing the proportion of bicep activities.
In this way, these findings can be communicated to different parties that help facilitate a more useful discussion. 
\looseness=-1

\subsection{User Interaction}
Our system provides various interactions besides the ones in \autoref{sec:bundle}. \looseness=-1

\paragraphem{Filtering.} Filtering exists in the file selection menu in Time Series View. 
Given more than 200 motion assessments, we provide an exclusive drop-down menu, 
that each drop-down only shows the available options filtered by the user's selection on other drop-downs. \looseness=-1

\paragraphem{Brushing.} The brushing interaction in the line chart illustrated in \autoref{fig:teaser}$\textbf{a}_{\textbf{3}}$ plays an important role in drilling down the final presentation
and outcome from the analysis so that users can achieve detail-on-demand in different stages of required actions (\textbf{R4}). 

\subsection{Implementation}
\autoref{fig:architecture} illustrates the system architecture. 
\systemname is a web-based application developed under Flask framework. 
The front-end widget functionality and plotting are achieved by Gridster and D3.js. 
The dataset is stored as Pandas Dataframe indexed with files and time steps,
and we store all the users' saved files in MongoDB.
We use NumPy for all data handling tasks. 
An important benefit of this approach is that we can vectorize all computations by treating the dataset as a matrix, 
allowing computations to be optimized in the low-level architecture. 
We deployed the back-end part into our server with 2GHz Intel Xeon E7-2850 CPU and 32GB memory. 
We achieve interactive speed in all calculations without precomputing any statistics and caching,
while our users from North America and Europe can work in their local machines and create, load and save their work without any installation.\looseness=-1

%% file: use-case.tex
\section{Case Studies}
\paragraph Our physicians found many interesting patterns and insights addressing clinical problems and furthermore challenges in the medical research domain. 
To better illustrate how they generated insights and collected evidence with \systemname, 
we present one case study focusing on addressing clinical research challenges, 
and one case study about helping physicians land clinical findings. \looseness=-1

\subsection{Trapezius Muscle Contribution to Shoulder Motion} 
\paragraph Our first case study summarizes how the physicians used the system 
to address the following much discussed clinical challenge\cite{bhatt2013middle,elhassan2016outcome,heuberer2015electromyographic}:
\newline

\textit{Do the upper trapezius (UT) and lower trapezius (LT) muscles 
have useful activities in the affected limb for shoulder motions?}
\newline

\paragraph It is an interesting question because the role of trapezius muscles in normal conditions is poorly understood, 
and physicians may consider the options to denerve these muscles, which they disconnect the nerves connecting to the muscles 
and reconnect the healthy nerves to the affected region, to provide a better distal ability. However, the effect of the loss of the trapezius muscle, 
while known to be debilitating in normal children, 
has to the best of our knowledge never been studied in children and adolescents with chronic obstetrical brachial plexus palsy. 
Therefore, using \systemname, our physicians addressed this problem with the comparative analysis of shoulder motions.

\paragraphem{Import, Inspect and Refine.} Our physicians imported the comparisons of the shoulder flexion motions between the same patients' limbs 
by selecting ``Shoulder Flexion'' in the muscle bundle comparison view. 
\revise{The first task before data analysis was to make sure all the data shown was clean (\textbf{T.1}).}
She discovered one of the motions' durations differed greatly between the limbs, 
therefore she opened the video by brushing the line chart to inspect the video to see what happened to the surprisingly long motion.
As the muscle activities of the affected limb in the extended duration did not seem to contribute to the task in the video, 
\revise{she use time series view to truncate the duration of the patient's motion.} 
Eventually, our physician acquired cleansed muscle bundle comparison results of all shoulder flexions \revise{ for further inspection (\textbf{T.2}).} 

\begin{figure}[tb]
    \includegraphics[width=\linewidth]{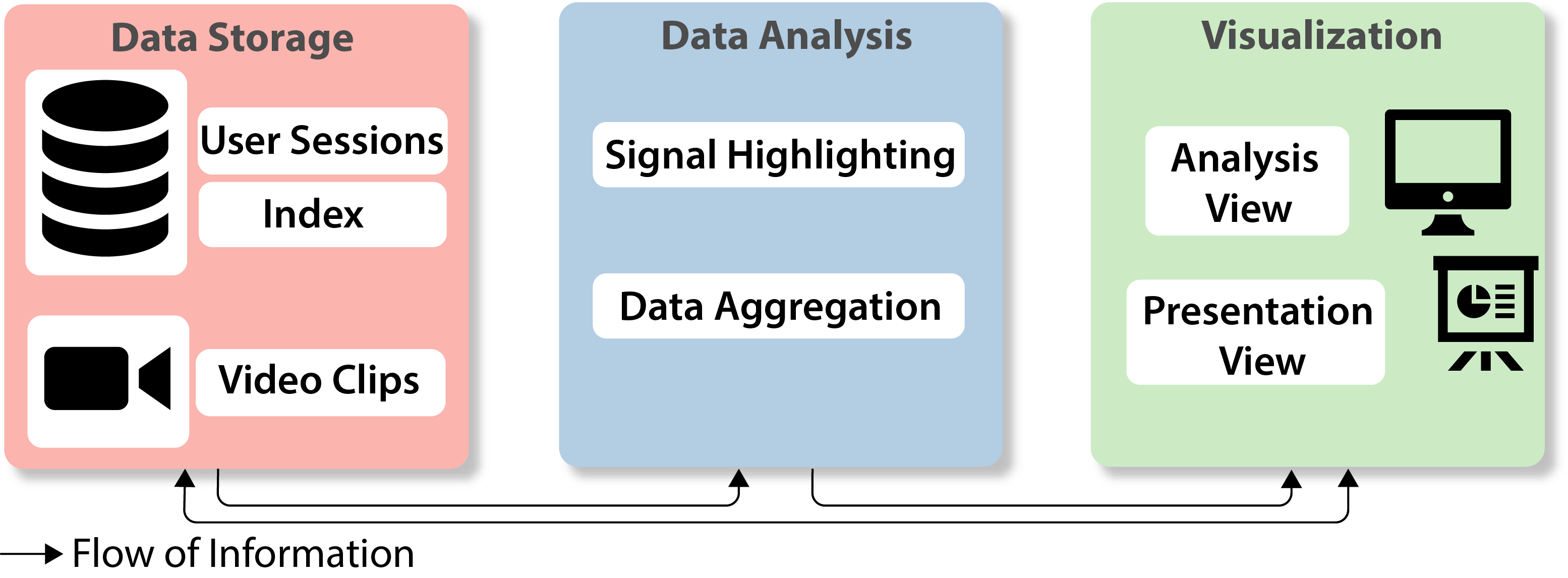}
    \vspace{-5mm}
    \caption{System architecture for data storage, modeling, and visualization.}
    \label{fig:architecture}
    \vspace{-3mm}
\end{figure}

\begin{figure*}
  \includegraphics[width=\linewidth]{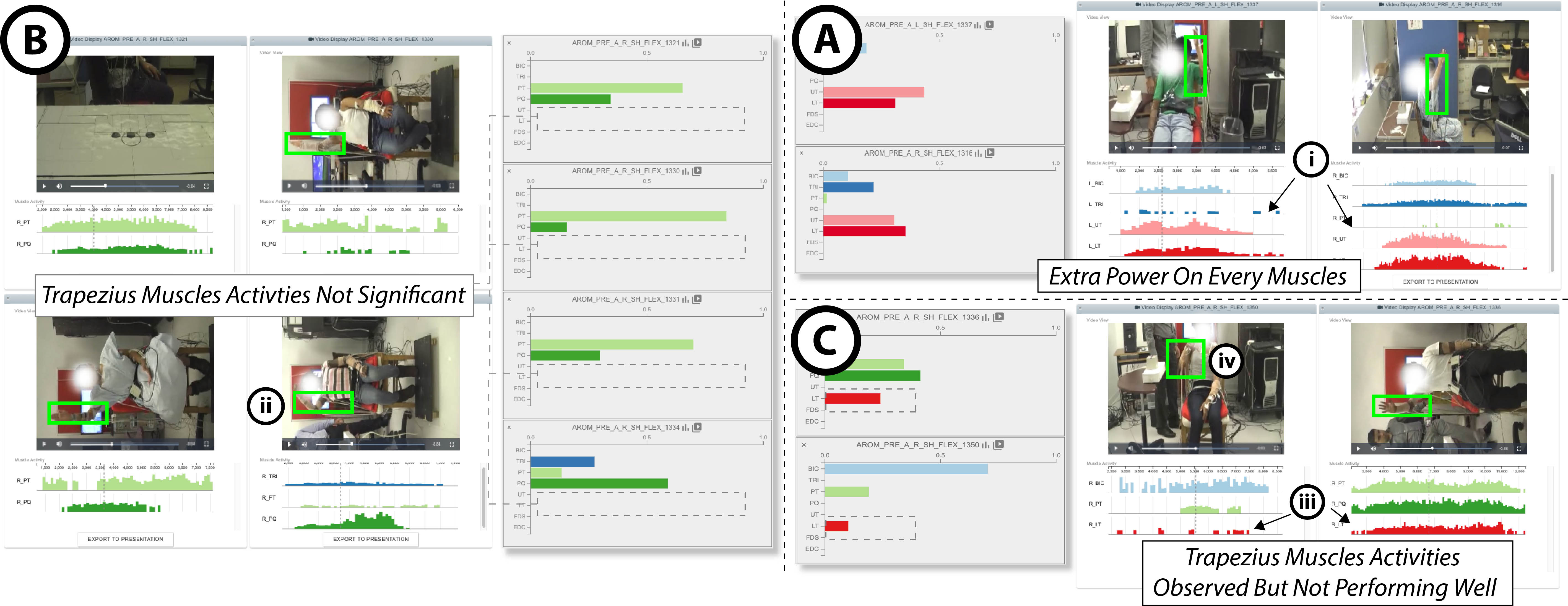}
  \vspace{-4mm}
  \caption{After using the analytic workflow shown in \autoref{fig:bundleComparison} on every patient to compare each affected and unaffected limbs,
our experts summarized three groups of patients to evaluate the usefulness of trapezius muscles (pink and red) on shoulder motions. 
\textbf{A}: Patients using extra powers on almost every muscle on the affected limbs shown in \clabel{i}; 
\textbf{B}: Patients with significantly stronger trapezius muscle activities on his unaffected limb shown in \clabel{ii}; 
\textbf{C}: Patients using more trapezius muscles shown in \clabel{iii}. 
Further inspections in \clabel{ii} suggested that profound debilitation was not significant among all patients in group \textbf{B}, 
while it could be observed among several patients in group \textbf{C}. 
These provide evidence to support spinal accessory denervation on patients with Obstetrical Brachial Plexus Palsy.}
  \label{fig:case1}
  \vspace{-4mm}
\end{figure*}

\paragraphem{Classify Patient Behavior Through Comparisons.} After a set of muscle bundle comparisons was generated,
our physicians \revise{used the visual highlighting filter to quickly remove similar and more insignificant muscle activties between the affected and unaffected limbs (\textbf{T.3}).} 
Given the stronger muscles between the limbs, they discovered that patients indeed behaved differently under three circumstances. 
Therefore they exported the analyzes of affected limbs from the comparison charts to video views (\autoref{fig:case1}) \revise{to generate a summary of comparison (\textbf{T.4}).}
To begin with, two of them (\autoref{fig:case1}\textbf{A}) emitted greater signals in almost every muscle.
Physicians discovered it by inspecting the analyzed comparisons that resulted in the video views of highlighted muscle activities in \autoref{fig:case1}\clabel{\textbf{i}}. 
The appearance of all muscles in the video views meant that the affected limbs 
fired nearly every muscle in greater magnitudes than the unaffected side. 
In this case, the patients' problems were only overshooting their muscles, indicating less severity. 
Furthermore, four patients clearly showed a lack of trapezius muscle activities in all shoulder motions (\autoref{fig:case1}\textbf{B}). 
It could be easily seen that only the pronator muscle activities remained (green muscles). 
However, our physicians were skeptical about their importance for the analysis, since they mainly served forearm's motions but not shoulder's. 
Last but not least, two patients were classified as having more activities on their trapezius muscles in their affected limbs, 
for which the low trapezius muscles could be significantly seen on the affected sides when highlighted muscles were inspected (\autoref{fig:case1}\clabel{iii}). \looseness=-1

\paragraphem{Collect Visual Evidence to Support Stance.} 
At this stage, our physicians would like to know \textit{how do patients behave with trapezius muscles?} and 
\textit{how do patients behave without trapezius muscles?} 
They, therefore, inspected the videos of patients' motions in the comparison view 
by brushing the time intervals with high limb displacement shown in the line chart \revise{to verify their hypotheses (\textbf{T.4}).} 
From the cut-scenes, our physicians were able to conclude the detected muscle activities quickly. 
For the patient group without more significant trapezius muscle activities, 
all of them performed the motions effectively (\autoref{fig:case1}\clabel{i}). 
Nonetheless, our physicians could see little debilitation on the patients when they perform shoulder activities. 
On the contrary, interestingly, our physicians could observe profound debilitation for some patients with stronger trapezius muscle activities 
on both affected and unaffected limbs. When they inspected the video cut scenes, 
they could see that the patient could not flex her shoulder or even raised her arm greater than around 45 degrees (\autoref{fig:case1}\clabel{iv}). 
Therefore they understood that the reason why the patient had a vigorous activity on her biceps 
was that she kept bending her forearm during flexing her shoulder. 
These visual clues, plus the facts from the physical outcomes, suggested that 
decreased trapezius muscle activities did not necessarily prohibit shoulder motions 
while having significant activities of them did not provide evidence that shoulder motions could be completed efficiently.
Thus our physicians concluded that, with particular regard to obstetrical brachial plexus cases, 
nerve transfer could be a seemingly attractive option compared to the alternative of exploring and grafting the C5 nerve root. 
The approach for nerve transfer from these muscles is more superficial and does not require the use of autologous nerve graft, 
as well as being a shorter distance to the muscle. 
They eventually arranged the results in the presentation view in three columns to convey such findings to other colleagues.

\paragraphem{Clinical Findings Besides Research Challenges.} 
Besides collecting evidence for answering research questions, 
our physicians found it helpful to classify patients into different categories, 
using the insights acquired by the bundle comparison chart. 
In identifying the three patterns, we can make conjectures about treatment options. 
For the first case where patients nearly overshot all of their muscles in their affected limb (\autoref{fig:case1}\textbf{A}), 
it demonstrated that they had a functioning cerebral pattern of activity. 
Our physicians believed that this subset may be better served by modulating duration, activation, deactivation, and coordination of muscle activity. 
For the second pattern where there were limited trapezius muscle activities (\autoref{fig:case1}\textbf{B}), 
patients still had adequate shoulder motion proofed by the evidence mentioned above. 
Such observation raised a question of how the motions had compensated using other muscles not attached by the sensors, 
thus further analysis could reveal what can be improved to go for a better function. 
In the last pattern (\autoref{fig:case1}\textbf{C}), 
having greater activations of trapezius muscle activities was not necessary to lead to good shoulder motion. 
Thus, our physicians conclude that other sources were responsible for the poor shoulder motion. 
In the future other shoulder muscles, 
such as deltoid, infraspinatus or supraspinatus muscles, 
should be added to the assessment.\looseness=-1

\subsection{Motion Analysis to Improve Clinical Evaluations}
\paragraph The second case study mainly focuses on applying \systemname to aid our physicians in addressing more general assessments when conducting clinical consultations.
Based on observation, our physician experts need to give a \textit{Narakas Classification}\cite{al2009narakas}, 
a grading system based on clinical observation to assess possible outcomes of children with obstetric brachial plexus palsy. 
Our physicians often had a question: ``\textit{Can we provide better anticipation before conducting clinical observation?}'' \looseness=-1

\begin{figure}[tb]
  \includegraphics[width=\linewidth]{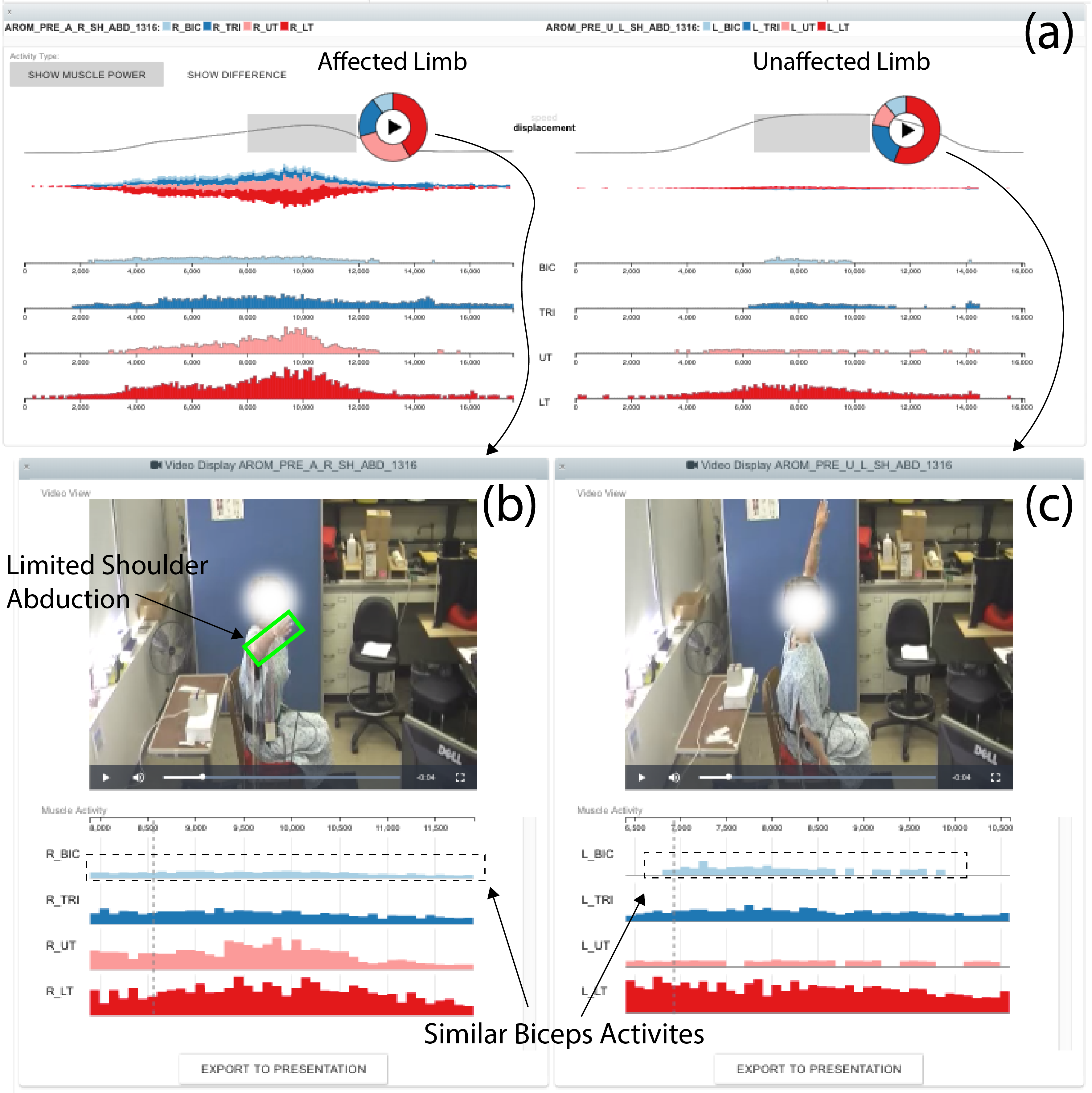}
  \caption{Usage scenario of using \systemname to anticipate clinical evaluation.  
  The Muscle Bundle Comparison View in (a) displays the muscle activities 
  from shoulder abduction conducted with the affected limb and the unaffected limb. 
  The visual clues of important muscles' activities combined with the difference of visible outcomes in (b) 
  and (c) provide explanations for medical classifications. \looseness=-1
  }
  \label{fig:case2}
  \vspace{-5mm}
\end{figure}

\paragraphem{Import Motions with Clinical Concerns.} Before going for a consultation, 
our expert first loaded the motions of shoulder abduction from the patient's affected and unaffected limbs to the time series view.
Then she discovered that the chart was distorted because the recording of one of the patient's motion was much longer, 
but there was no sign of vigorous movement after the middle of the assessment. 
Also, the unaffected limb had a strong pronator (PQ and PT) and flexor digitorum superficialis (FDS) muscle activities. 
She concluded that they were not helpful in shoulder motions because PQ and PT muscles were responsible for forearm movement and FDS was for fingers. 
Therefore, she shortened both activities to around 16 seconds 
and removed the inspection of other muscles except biceps, triceps and trapezius muscles \revise{for a more focused inspection on the clinical purpose (\textbf{T.1}).} . 
Now, the results in the muscle comparison view from importing the cleansed time series became much clearer 
since there were only four muscles in the bundle comparison chart (\autoref{fig:case2}(a)).
\revise{This allowed the physician to conduct individual limb analysis (\textbf{T.2}).}

\paragraphem{Acquire Facts Before Consultation.} 
Our experts then inspected the coordination of muscles within one limb. 
After filtering some irrelevant portion with the aid of the displacement function line chart \revise{(\textbf{T.3})}, 
it appeared that all muscles were active when the patient conducted shoulder abduction  
with similar total distribution among the limbs. 
However, \revise{to verify the findings using information apart from the muscle activities (\textbf{T.4})}, 
our physician inspected the physical outcome of the patient in the video clip (\autoref{fig:case2}(b)) 
and observed that the patient did not extend her shoulder as high as her unaffected limb in \autoref{fig:case2}(c). 
It could be seen that the affected limb had a similar output of muscle activities without a desirable outcome. 
While whether the functions of trapezius muscle were important was still unclear, 
a lack of effectiveness of biceps activity with a limited degree of abduction shown in the video indicated a Narakas I characteristics. 
While our physicians could make such a conclusion after inspecting the data, whether the patient would fall into Narakas II depended on her wrist movement. 
Nevertheless, they acquired a brief and speedy inspection of muscle behavior with muscle bundle comparisons and video inspection in \systemname.

\paragraphem{Hypotheses to Current Assessment Methods.} 
Our physicians checked their previous documentation, 
and found that the patient belonged to the ``low functioning'' group, 
meaning the patient had ``\textit{a range of active motion in shoulder flexion and abduction $<$ 90 degrees on the affected limb}''.
\revise{Beforehand, patients were first classified into low and high functioning groups
according to their degrees of shoulder movement, then physicians tried to see if there were
any differences in muscle activity between the two groups.}
Yet, after checking the muscle activities and physical outcome of the patient, 
our physicians were not convinced that the patient was low functioning 
since there were still some biceps activities and the movement did not look that minimal. 
Our physicians thus suspected the classification of patients by such methodology. 
With \systemname, they tended to rely more on a visual inspection 
and used the patterns mentioned with the previous case to classify patients, 
as it looked more convincing and the visual facts were more holistic and presentable.  

%% file: conc.tex
\subsection{Expert Interview}
\label{section:evaluation}
Our physicians were impressed with the designs. 
They commented that the arrangement of placing every patients' performance one by one with the same design made the analysis extremely convenient and efficient. 
Also, the ability to offset every similar muscle signals made the analysis easy to proceed. 
They emphasized that these conveniences were crucial in the diagnosis process.
Physicians often did not want to spend too much time on acquiring information since they went through different kinds of diagnoses besides data analysis.
Therefore, they regarded visual analytics demonstrated in the system as an efficient and explainable data analysis tool.

We also identify a usability issue in the system. Our physicians raised an issue of the occasional confusion of color encodings on the muscles.
Thus, the system always provides labeling besides the muscles in all the views, so that when users encounter different color bars, 
they can always refer to the names of the muscles shown beside the charts. \looseness=-1

\section{Discussion}
\label{section:lessons}

\paragraphem{Lessons Learned.} Working with physicians, we learned a new perspective of valuing the importance of visual analytics when users need to work \textit{on} the data. 
The visual analytics mantra ``overview first, zoom and filter, then details-on-demand'' was built on the general goal of \textit{generating insights} \cite{saraiya2006insight}. 
Yet, similar to many clinical workflows, our situation was about \textit{insights through iterated inspections}, 
that required lots of inevitable trial and error processes.
Therefore, our physicians emphasized the greatest benefit of our system was about shortening their cycles of analyzing each patient \revise{from clicking multiple muscles to using a slider to filter the most insignificant muscles within one drag}.
Users might not receive any insights during the analysis of an individual patient, 
but what makes the analysis useful is the ability to quickly extract the useful facts in a single iteration. 
As a result, users could acquire an overview of useful information that eventually produces an insightful conclusion.

Another lesson learned was an enforced impression on the importance of verifiable visualization techniques.
Our validity of using sliders with the visual highlighting method to highlight important muscles 
relied much on the presence of original muscle signals encoded with strokes to keep the original data (\autoref{fig:bundleComparison}(d)).
When it came to clinical decisions that are prone to false information, 
the effectiveness of visualization lied in the ability to verify but not solely on the ability to generate insights.
It is crucial to prevent ourselves from falling one of the \textit{abstraction threats} mentioned in the
nested models for visualization design and validation: ``operations and data types do not solve the characterized problems''.
In our situation, our mistakes at the beginning were that we tried to aggregate the muscle coordination for more intuitive visuals like projections or clustering,
but it was some iterations later that we figured out the importance of avoiding any automated feature extraction when the data itself was ambiguous.


\revise{\paragraphem{Evaluation.} The effectiveness of our system is mainly evaluated by the clinical impacts and medical findings made by our domain experts. Our limitation lies in a lack of concrete numerical values such as task completion time or A/B testing to demonstrate our effectiveness. We are currently working on using our system to conduct more in-depth medical related user experiments.}

\paragraphem{Scalability.} Our system can handle feature extraction and comparisons of multivariate temporal data in an interactive speed, 
but it does not support visual comparisons of a large number of signals since we use color to distinguish them. 
Though it is unrealistic to attach hundreds of sensors to the patient, 
the perception of noticeable differences between colored signals will diminish when there are more than 12 lines\cite{munzner2014visualization}. 
If there is a need for comparing a muscle bundle of more than 12 temporal attributes, 
we will investigate more on prioritizing time series data for visualization. 
Rong \etal have investigated the topic of prioritizing deviation of univariate temporal data\cite{rong2017asap}, 
and we believe that establishing an attention aware strategy of prioritizing multi-attribute temporal data will be a promising direction.

%


\paragraphem{Application Domain.}  Although our work is primarily designed for EMG signal bundle comparison, 
it can be easily adapted for other similar problems, such as adult motion analysis or sports injury analysis. 
In these problems, we can apply similar techniques to compare the different behavior between different body-parts in respective motions. 
As we broadcast \systemname to a greater amount of physicians of the Pediatric Upper Extremity Motion Analysis Program, 
usability becomes an important aspect of the software development in the future, and we need a formal user study to test its usability.
We hope to introduce our system to a wider audience in the domain of motor recovery research so that the system will be more generalized for 
various kinds of recovery studies.

\section{Conclusion}
\label{section:conclusion}
\paragraph In this paper, we presented \systemname, 
a visual analytics system that allows users to interactively compare muscle bundles' activity from patients under obstetrical brachial plexus injuries. 
\systemname proposes techniques for efficient analysis of muscle signal bundles and
 integrates different sources of heterogeneous data into consistent and coordinated views, 
 thus aids physicians to understand nerve coordinations under obstetrical brachial plexus palsy. \looseness=-1